\documentstyle[12pt]{article}
   
\def\be{\begin{equation}}
\def\ee{\end{equation}}
\def\ben{\begin{displaymath}}
\def\een{\end{displaymath}}
\def\ba{\begin{array}{c}}
\def\ea{\end{array}}
\begin{document}

\titlepage
\vspace*{2cm}

\begin{center}{\Large \bf
An asymptotic intertwining of the undelayed and delayed Fibonacci
numbers
 }
\end{center}

\vspace{10mm}

\begin{center}
Miloslav Znojil
\vspace{3mm}

\'{U}stav jadern\'e fyziky AV \v{C}R, 250 68 \v{R}e\v{z},
Czech Republic\\

e-mail: znojil @ ujf.cas.cz

\end{center}

\vspace{5mm}

\section*{Abstract}

The list of properties of Fibonacci numbers $F_n$ (with
multifaceted relevance in physics) is complemented by an empirical
observation that in combination with the ``next" family of the
``delayed Fibonacci" numbers $G_m$ (called, for convenience,
``Gibonacci numbers" here), both sets exhibit certain remarkable
and fairly unexpected asymptotic mutual-bracketing properties.

\vspace{5mm}

 PACS 02.50.Kd; 02.70.Rw; 02.90.+p;  05.50.+q

\newpage

\section{Introduction}

Our forthcoming considerations were inspired by the decoration of
the northmost subway station ``Hole\v{s}ovice" in Prague. Any
passenger who waits there for the train to the city may wonder why
the architects ornamented the walls by the regularly repeated
lines of 2, 2, 3, 4, 5, and 6 tiles. In this sense, our present
paper comes only too late when suggesting a replacement of such a
pattern by an inessentially modified sextet of the integer numbers
$G_3=2$, $G_4=2$, $G_5=3$, $G_6=4$, $G_7=5$ and $G_8=7$.

\section{Fibonacci-type recurrences with zero- and one-step
delay}

Our deeper motivation stems from the fact that the above-mentioned
segment of a sequence of $G_n$ may be understood as one of the
most natural ``delayed" modifications of the famous Fibonacci
numbers \cite{Fibonacci}. The latter sequence with conventional
denotation $F(n+1) \equiv F_{n}$ \cite{definition} is defined by
the well known and extremely elementary recurrent relations
 \be
 F_{n}=F_{n-1} + F_{n-2}, \ \ \ \ \ \ \ \
 n = 2,3 \ldots\,,
 \ \ \ \ \ \ \ \ \ \ F_0=F_1=1\,.
 \label{nedelay}
 \ee
Similarly, their ``delayed" version may be generated by the very
similar ``next-to-Fibonacci" recurrences with a one-step delay,
 \be
 G_{n}=G_{n-2} + G_{n-3}, \ \ \ \ \ \ \ \
 n = 3, 4, \ldots\,,
 \ \ \ \ \ \ \ \ \ \ G_0=G_1=G_2=1\,.
 \label{delay}
 \ee
In our present note we intend to present a few arguments showing
why both the sequences $F_n$ and  $G_n$ might be considered
comparably interesting.

Firstly, let us remind the reader that both these sequences are
virtually equally easy to construct. Both difference equations
(\ref{nedelay}) and (\ref{delay}) for $F_n$ and $G_m$ have
constant coefficients and may be analyzed by the similar standard
ansatzs, viz.,
 \be
 F_n=
 a\,\left [ \eta_{(a)} \right ]^n+
 b\,\left [\eta_{(b)}\right ]^n
 \label{ansFi}
 \ee
and
 \be
 G_n=
 A\,\left [ \varrho_{(A)} \right ]^n+
 B\,\left [ \varrho_{(B)} \right ]^n+
 C\,\left [ \varrho_{(C)} \right ]^n\,
 \label{ansGu}
 \ee
respectively. This procedure leads to the similar implicit
definitions of the necessary quotients.

\subsection{Fibonacci sequences \label{2} }

In the former case which corresponds to the usual Fibonacci
numbers $F_n$ the implicit definition of the quotient is slightly
simpler,
 \be
 \left [ \eta_{(a,b)} \right ]^2=
 \eta_{(a,b)}+1\,.
 \label{latter}
 \ee
This equation may be immediately assigned the following well known
and complete explicit solution,
 \be
 \eta_{(a)} = \frac{1+\sqrt{5}}{2}
 \,,
 \ \ \ \ \ \
 \eta_{(b)} = \frac{1-\sqrt{5}}{2}= -\left (\eta_{(a)}-1
 \right )
\,.
 \ee
In it, the former item coincides with the famous ``golden mean"
value.

\subsection{``Gibonacci" sequences  \label{2.1}}

There is really no reason why one should be afraid of searching
for the similar explicit solutions in the delayed case with the
similar implicit definition
 \be
 \left [ \varrho_{(A,B,C)} \right ]^3=
 \varrho_{(A,B,C)}+1\,
 \label{former}
 \ee
of the quotients. Routinely, one arrives at just one real and
positive root
 \be
 \varrho_{(A)} =D_+ +D_-
 \ee
where we abbreviated
 \ben
  D_{+} =
 \left ( \frac{1+ x_0}{2}\right )^{1/3} \approx
 0.98699\,,
 \ \ \ \ \ \ \ \ \
 D_{-} =
 \left ( \frac{1- x_0}{2}\right )^{1/3} \approx
 0.3377\,
 \een
and where
 \ben
 \ \ \ \ \ \ \ \ \
  x_0=\sqrt{\frac{23}{27}} \approx 0.9229582\,.
 \een
The other two complex conjugate roots possess the equally compact
representation
 \be
 \varrho_{(B)} =
 \left [ \varrho_{(C)}\right ]^* =e^{i\,\varphi}D_+ +
 e^{-i\,\varphi}D_-\ \left (=
 -\frac{1}{2}\,\varrho_{(A)}
 + i\,
 \frac{\sqrt{3}}{2}\,
 \varrho_{(D)} \right )
 \ee
where $\varphi=2\pi/3$ and $ \varrho_{(D)} =D_+ -D_- \approx
0.649264 $.

\subsection{Specific initial boundary conditions \label{2.2}}

Once we require the compatibility of formula (\ref{ansFi}) with
the Fibonacci's boundary conditions $F_0=F_1=1$, we have to
extract the corresponding values of $a$ and $b$ from the two
equations
  \be
1 = a+b
 =
 a\, \frac{1+\sqrt{5}}{2}
+
 b\,\frac{1-\sqrt{5}}{2}\,.
 \label{HansFi}
 \ee
This gives the well known explicit formula for Fibonacci numbers,
 \be
 F_n=\frac{1}{\sqrt{5}}\,
\left [
 \left ( \frac{1+\sqrt{5}}{2}  \right )^{n+1}-
 \left ( \frac{1-\sqrt{5}}{2} \right )^{n+1}\right ] \,.
 \label{fiansFi}
 \ee
In the similar manner, the delayed-Fibonacci general solution
(\ref{ansGu}) complemented by the boundary conditions
$G_0=G_1=G_2=1$ will define our present Gibonacci numbers $G_m$.
First of all we  must find the correct values of the coefficients
in eq. (\ref{ansGu}) by solving the the triplet of the linear
equations
 \be
 1=A+B+C=
 A\, \varrho_{(A)} +
 B\, \varrho_{(B)} +
 C\, \varrho_{(C)} =
 A\,\left [ \varrho_{(A)} \right ]^2+
 B\,\left [ \varrho_{(B)} \right ]^2+
 C\,\left [ \varrho_{(C)} \right ]^2\,.
 \label{ansGube}
 \ee
Its numerical solution gives the approximate form of the result,
 \be
 \ba
 A =
 0.72212441830311284114\ldots,
  \\
 B =
 0.13893779084844357942\ldots - i \cdot 0.20225012409895253966 \ldots,
 \\    C =
 0.13893779084844357942\ldots + i \cdot 0.20225012409895253966 \ldots\,.
 \ea
 \label{check}
 \ee
Now, the same result will be derived in closed form, completing in
this way the analogy with the previous ``non-delayed" formula
(\ref{fiansFi}).

Firstly, we suppress all temptations to use a computerized
symbolic manipulations and put $B=K-i\,L = C^*$. The first item in
eq. (\ref{ansGube}) then offers the direct elimination of $A = 1 -
2K$. Moreover, after an abbreviation
 \be
 1+2\,K\,( \cos \varphi - 1 ) + 2\,L\,\sin \varphi = \Sigma\,,\ \
 \ \ \ \ \ \
 1+2\,K\,( \cos \varphi - 1 ) -2\,L\,\sin \varphi = \Delta\,,
 \label{hoh}
 \ee
we may re-write the remaining two lines of eq. (\ref{ansGube}) in
a particularly friendly form,
 \be
 \ba
 \Sigma \,D_+\,+ \Delta \,D_-\,= 1\\
 3\,\Sigma \,D_-^2+ 3\,\Delta \,D_+^2= 1\,.
 \ea
 \ee
This two-by-two matrix equation is readily solvable,
 \ben
 \Sigma=
 \frac{1}{3x_0}\,\left (
 3\,D_+^2-{D_-}
 \right )\,,\ \ \ \ \ \ \ \ \ \ \
 \Delta=
 \frac{1}{3x_0}\,\left (
 -3\,D_-^2+{D_+}
 \right )\,.
 \een
The final backward insertion in eq. (\ref{hoh}) is trivial and
gives the final answer,
 \ben
 6\,K= 2 - \frac{1}{3x_0}
 \left (D_+ -D_-
 \right )
 \left (
 3D_+ +3D_- + 1
 \right )\,,
 \een
 \ben
 2\,\sqrt{3}\,L =
 \frac{1}{3x_0}
 \left (
 3D_+^2+3D_-^2 - D_+ -D_-
 \right )\,
 \een
which is compatible with its numerical check (\ref{check}).

\section{Intertwining behavior of the Fibonacci-type
sequences
 \label{3}}

\subsection{Asymptotics of $F_n$ and $G_m$ at the large indices
 \label{3.1}}

As long as we observe that
 $
 |\eta_{(a)}| = 1.618\ldots > 1$ and $
 | \eta_{(b)}| = 0.618\ldots < 1
 $
while
 $
 |\varrho_{(A)} |
 =
 1.3247\ldots > 1$ and $
 |\varrho_{(B)} |=
 |\varrho_{(C)} |\,
 =
 |-0.662\ldots
 \pm i\,0.562\ldots
 |$, i.e., $|\varrho_{(B)} |
 =
.8688\ldots < 1$, we may conclude that irrespectively of the
initial boundary conditions the solutions of both the difference
equations (\ref{delay}) and (\ref{nedelay}) will always exhibit a
similar asymptotic behavior. The reason is that both $F_n$ and
$G_m$ are dominated by the single power-law term, i.e.,
 \be
 F_n=
 a\,\left [ \eta_{(a)} \right ]^n+
 + {\cal O}
 \left \{ \left | \eta_{(b)} \right |^n
 \right \}, \ \ \ \ \ \ \ \ \ \ n \gg 1\,
 \ee
while
 \be
 G_m=
 A\,\left [ \varrho_{(A)} \right ]^m+
{\cal O}
 \left \{ \left | \varrho_{(B)} \right |^m
 \right \}, \ \ \ \ \ \ \ \ \ \ m \gg 1\,.
 \ee
From these relations we may deduce that the size of the numbers
$F_N$ and $G_M$ cannot remain comparable unless the indices $N \gg
1$ and $M \gg 1$ obey the following rule,
 \be
 \frac{N}{M} \approx \frac{\ln \varrho_{(A)}}{\ln \eta_{(a)}}
 =
 0.584\ldots\,.
 \label{ratio}
 \ee
This means that the replacement of the Fibonacci recurrences
(\ref{nedelay}) by their one-step-delayed modification
(\ref{delay}) slows down the asymptotic growth of the new
sequence, $G_M \approx F_{entier\{0.584\,M\} }\ $ at $ M \gg 1$.

\subsection{Inequalities between $F_n$ and $G_m$ at the finite
indices
 \label{3.2}}

An identity
 \be
 \frac{\ln \varrho_{(A)}}{\ln \eta_{(a)}}
 =
 0.58435715765740408667...=
 \frac{7}{12} + \frac{3}{10^3}\,\left (\frac{7}{12}
 \right )^2+ \frac{3}{10^6} -
 \frac{9}{10^9} +{\cal O}
 \left (
 \frac{1}{10^{11}}
 \right )\,
 \label{jedenact}
 \ee
complements eq. (\ref{ratio}) and indicates that the ratio $7/12$
may play a key role in our present analysis. Indeed, once we
tentatively re-index all the very large Fibonacci numbers,
 \be
 F_N = f(j,k), \ \ \ \ \ \ \
 N = N(j,k) = 7\,j + k,\ \ \ \ \ \ \
  \ \ \ \ \ \ \ 0 \leq k \leq 6,
  \label{numF}
   \ee
and once we re-write their delayed-generated alternative in the
similar form,
 \be
 G_M = g(J,K), \ \ \ \ \ \ \
 M = M(J,K) = 12\,J + K, \ \ \ \ \ \ \ \
 0 \leq K \leq 11,
  \label{numG}
 \ee
we may re-interpret the above ``asymptotic comparability rule"
(\ref{ratio}) as a requirement
 \be
 \frac{7\,j + k}{12\,J + K} \approx
  \frac{7}{12}\,,\ \ \ \ \ \ \ \ \ \
  j ,J \gg 1\,.
  \label{dvanact}
 \ee
In the other words, we achieve the approximative asymptotic
coincidence of $G_M = g_{J,K}$ with $F_N = f_{j,k}$ if and only if
the new auxiliary indices $j $ and $J$ do not differ too much.

\subsection{Inequalities between $F_n$ and $G_m$ at the small
indices
 \label{3.3}}

{\it A priori}, there is no reason to believe that the similar
rule could be extended to the domain of the small indices.
Nevertheless, we may take the union set of {\em all} the numbers
in eqs. (\ref{numF}) and (\ref{numG}) and {\em order} this family
in a way starting at the very first subscripts. In this way the
first seven lines of inequalities are revealed,
 \be
 \begin{array}{l}
 g(0,-1) \ (= 0)\, < f(0,0) \ (=1)\, \leq g(0,0)\ (=1),\\
 g(0,1) \ (= 1)\, \leq f(0,1) \ (=1)\, \leq g(0,2)\ (=1),\\
 g(0,3) \ (= 2)\, \leq f(0,2) \ (=2)\, \leq g(0,4)\ (=2),\\
 g(0,5) \ (= 3)\, \leq f(0,3) \ (=3)\, < g(0,6)\ (=4),\\
 g(0,6) \ (= 4)\, < f(0,4) \ (=5)\, \leq g(0,7)\ (=5),\\
 g(0,8) \ (= 7)\, < f(0,5) \ (=8)\, < g(0,9)\ (=9),\\
 g(0,10) \ (= 12)\, < f(0,6) \ (=13)\, < g(0,11)\ (=16)\,.
 \ea
 \label{naught}
 \ee
Encouraged by the smoothness of this pattern we may verify, with
an utterly unexpected success, the existence and validity of its
next-step continuation
 \be
 \begin{array}{l}
 g(0,11) \ (= 16)\, < f(1,0) \ (=21)\, \leq g(1,0)\ (=21),\\
 g(1,1) \ (= 28)\, < f(1,1) \ (=34)\, < g(1,2)\ (=37),\\
 g(1,3) \ (= 49)\, < f(1,2) \ (=55)\, < g(1,4)\ (=65),\\
 g(1,5) \ (= 86)\, < f(1,3) \ (=89)\, < g(1,6)\ (=114),\\
 g(1,6) \ (= 114)\, < f(1,4) \ (=144)\, < g(1,7)\ (=151),\\
 g(1,8) \ (= 200)\, < f(1,5) \ (=233)\, < g(1,9)\ (=265),\\
 g(1,10) \ (= 351)\, < f(1,6) \ (=377)\, < g(1,11)\ (=465)\,.
 \ea
 \label{draught}
 \ee
Now, there comes one of our main empirical observations. Against
all odds, the same scheme works during unexpectedly many
iterations numbered by the integer $K$ which appears as the first
argument in the functions $g(K,*)$ and $f(K,*)$ and which was
equal to zero in eq. (\ref{naught}) and to one in the subsequent
set of the fourteen inequalities (\ref{draught}). We arrive at the
formidably extensive set of the empirical inequalities
 \be
 \begin{array}{ll}
 g(K-1,11) \ < f(K,0) \ < g(K,0)\ ,& \ \ K < 48 \\
 g(K,1) \ < f(K,1) \  < g(K,2)\ ,& \ \  K < 35\\
 g(K,3) \ < f(K,2) \  < g(K,4)\ ,& \ \  K < 21\\
 g(K,5) \  < f(K,3) \  < g(K,6)\ ,& \ \  K < 7\\
 g(K,6) \  < f(K,4) \  < g(K,7)\ ,&  \ \ K < 41\\
 g(K,8) \  < f(K,5) \  < g(K,9)\ ,& \ \  K < 27\\
 g(K,10) \  < f(K,6) \  < g(K,11)\ ,& \ \  K < 14\,.
 \ea
 \label{nhdraught}
 \ee
The existence and structure of the upper limits of their validity
reminds us of the fact that the next correction to eq.
(\ref{dvanact}) (given in eq. (\ref{jedenact})) does not vanish.
This means that the range of the allowed $K$ in eq.
(\ref{nhdraught}) cannot be unlimited.

\subsection{Inequalities between $F_n$ and $G_m$ at the growing
indices
 \label{3.4}}

It is remarkable that the inequalities (\ref{nhdraught}) are
violated {\em so extremely slowly} and in such an unbelievably
regular manner. This is one of the main consequences of the
smallness of the absolute value of the second correction $\approx
0.003\cdot (7/12)^2$ to the rule $N/M=7/12$ because it will enable
us to extend our inequality pattern beyond its limits listed in
eq. (\ref{nhdraught}).

As long as the next correction to the asymptotic comparability
rule $N/M=7/12$ is positive, we know {\em in advance} that the
middle terms $f(K,J)$ in eq. (\ref{nhdraught}) will all move to
the left and violate their lower estimates at a critical $K$. Due
to this expectation (confirmed by the explicit calculations in
MAPLE), the next-step bracketing law acquires the mere following
shifted form
 \be
 \begin{array}{ll}
 g(K-1,10) \ < f(K,0) \ < g(K-1,11)\ ,& \ \ 48 \leq K < 96 \\
 g(K,0) \ < f(K,1) \  < g(K,1)\ ,& \ \  35 \leq K < 82\\
 g(K,2) \ < f(K,2) \  < g(K,3)\ ,& \ \  21 \leq K < 70\\
 g(K,4) \  < f(K,3) \  < g(K,5)\ ,& \ \  7 \leq K < 55\\
 g(K,5) \  < f(K,4) \  < g(K,6)\ ,&  \ \ 41 \leq K < 89\\
 g(K,7) \  < f(K,5) \  < g(K,8)\ ,& \ \  27 \leq K < 75 \\
 g(K,9) \  < f(K,6) \  < g(K,10)\ ,& \ \  14 \leq K < 61\,.
 \ea
 \label{htnhdraught}
 \ee
Etc. One should notice that due to the fact that all the $K$'s are
already very large, all the separate intervals of validity of the
innovated rule (\ref{htnhdraught}) are perceivably longer than
their respective predecessors in eq.(\ref{nhdraught}). This seems
to indicate a general tendency, the detailed analysis of which
would require much more space than available here.

\section{Outlook \label{4}}

\subsection{Towards combinatorial applications \label{4.2}}

Fibonacci recurrences (\ref{nedelay}) (without any delay) are
extremely popular and find (perhaps, unexpectedly) numerous
practical applications. {\it Pars pro toto}, Fibonacci numbers
$F_n$ occurred recently as a sequence which numbers all the
possible re-arrangements of the Born-Lanczos expansions of the
scattering amplitudes in quantum mechanics \cite{Lippschw}. The
mathematical essence of this particular application lies, in a way
illustrated by Table 1, in the following combinatorial
representation of the Fibonacci numbers,
 \be
 F_k=
 \left (
 \ba
 k\\
 0
 \ea
 \right )+
 \left (
 \ba
 k-1\\
 1
 \ea
 \right )+
 \left (
 \ba
 k-2\\
 2
 \ea
 \right )+ \ldots\ .
 \label{fiboco}
 \ee
One may feel inspired to re-interpret the latter property as a
definition. In the next step one then could modify this type of
definition, obtaining the following ``higher" Fibonacci numbers,
 \be
 F^{(2)}_k=
 \sum_{j\geq 0{\rm \ while\ }3j\leq k}
 \left (
 \ba
 k-2j\\
 j
 \ea
 \right )
 =
\left (
 \ba
 k\\
 0
 \ea
 \right )
 +
 \left (
 \ba
 k-2\\
 1
 \ea
 \right )+
 \left (
 \ba
 k-4\\
 2
 \ea
 \right )+ \ldots\
 \label{h1fiboco}
 \ee
(cf. Table 2) or
 \be
 F^{(3)}_k=
 \sum_{j\geq 0{\rm \ while\ }4j\leq k}
 \left (
 \ba
 k-3j\\
 j
 \ea
 \right )
 =
 \left (
 \ba
 k\\
 0
 \ea
 \right )+
 \left (
 \ba
 k-3\\
 1
 \ea
 \right )+
 \left (
 \ba
 k-6\\
 2
 \ea
 \right )+ \ldots\
 \label{h2fiboco}
 \ee
(cf. Table 3) etc. From this background, one can always return to
the recurrent approach, discovering that it starts from a general
initial $\ell-$plet of values
 \be
 F_k^{(\ell)} = 1, \ \ \ \ \ \
 k = 0, 1, \ldots, \ell-1,\ell,
 \ \ \ \ \ \ \ \ \ \ell = 1, 2, \ldots\,.
 \ee
Moreover, the old combinatorial definitions (\ref{h1fiboco}) or
(\ref{h2fiboco}) (etc) become replaced by the corresponding
Fibonnaci-type new three-term-like recurrences
 \be
 F_{n}^{(2)}=F_{n-1}^{(2)} + F_{n-3}^{(2)}, \ \ \ \ \ \ \ \
 n = 3, 4, \ldots\,
 \label{praltdelay}
 \ee
or
 \be
 F_{n}^{(3)}=F_{n-1}^{(3)} + F_{n-4}^{(3)}, \ \ \ \ \ \ \ \
 n = 4, 5, \ldots\,
 \label{dfaltdelay}
 \ee
etc. In this way we would obtain a new series of generalized
Fibonacci numbers. The analysis of the influence of the delay in
the underlying recurrences lies already beyond the scope of our
present short communication but it is worth mentioning that it
would proceed precisely along the lines applied here to the
simplest delayed case~(\ref{delay}).

\subsection{An appeal of linearity \label{4.22}}

An apparent ambiguity of an initialization of the delayed
Fibonacci recurrences (\ref{delay}) is just fictitious. Indeed,
although a more general choice of the initialization appears
admissible,
 \be
G_0(a)=G_2(a)=1, \ \ \ \ G_1(a)=a \in (-\infty,\infty)\,
\label{newinic}
 \ee
the question has an elementary answer since the innovated
initialization (\ref{newinic}) merely produces the sequence with
elements $G_3(a)=1+a$, $G_4(a)=1+a$, $G_5(a)=2+a$, $G_6(a)=2+2a$,
$G_7(a)=3+2a$, $G_8(a)=4+3a$ etc. We immediately see that we have
 \be
 G_n(a) = G_{n-2}(1) + a\,G_{n-3}(1)\,
\label{newiw}
 \ee
so that all the variations of $a \neq 1$ do not induce any real
gain in generality.

The rule of this type may be understood as one of the
manifestations of the linearity of our present example. This
property opens a path towards applications of the similar models
in statistics where certain generalized Fibonacci numbers proved
related to the close-packed dimers on non-orientable surfaces
\cite{LU}, to the local temperature distribution on quasiperiodic
chains \cite{Mas}, to the existence of fractional statistics in
quantum gases of quasiparticles \cite{Rac}, to the statistics born
by the stacking of squares on a staircase \cite{Tur} and to the
electron and phonon excitations in quasicrystals \cite{Vel}.

Closely related use of the linearity of the three-term
Fibonacci-like recurrences helped, in \cite{Sir}, to introduce
randomness directly in the coefficients $a$ in a way resembling
eq. (\ref{newiw}), with possible impact ranging from the modelling
of chaos (e.g., in quantum gate networks and quantum Turing
machines with the Turing head controlled by a Fibonacci-like
sequence of rotation angles \cite{Kim}) till explicit models of
the critical level-spacing distributions, with eigenvectors lying
between extended and localized \cite{Kat}. Last but not least, a
strong appeal of all these models (which may all be interpreted as
various discretized versions of Schrödinger equations) lies in
their generic non-Hermiticity (cf., e.g., \cite{Der}) which became
subject to an intensive study recently (cf., e.g., \cite{BB} or
all papers in the dedicated issue \cite{Cz}).

\subsection{Numerical ``miracles" and open questions
 \label{4.1}}

An exceptional role of our {\em most elementary} modification
(\ref{delay}) of the {\em most popular} Fibonacci's recurrences
has been illustrated here with a particular emphasis on some of
the {\em most interesting numerical features} of the ``Gibonacci"
sequence $G_n$. In this sense, our key message has been based, in
essence, on the remarkably quick convergence of the asymptotic
series (\ref{jedenact}).

We did not pay attention to all the properties of $G_m$ of the
similar numerical type. For example, we did not throw any light on
an alternative numerical relation between our asymptotic quotient
$ \varrho_{(A)}$ and the golden mean $\eta_{(a)}$ which is based
on the evaluation of the logarithm of their ratio,
 \be
 \ln \left (\frac{ \varrho_{(A)}}{ \eta_{(a)}} \right ) =
 0.20001225073664160\ldots\,.
 \label{motiv}
 \ee
We see that it has an exceptional form (with several zero digits
in it) as well as a remarkably compact approximate representations
with higher precision, e.g.,
 \be
 \frac{1}{5} + \frac{1}{8\cdot 10^4} - \frac{1}{4\cdot 10^6}
 +\frac{2}{27\cdot 10^8} -\frac{41}{10^{13}} =
 0.20001225073664074\ldots\,.
 \ee
The smallness of the subsequent corrections as well as the use of
the natural logarithm in eq. (\ref{motiv}) do not have in fact
{\em any} natural explanation. In the other words, the
comparatively high reliability of the estimate
 \be
  \varrho_{(A)}^5 = e\,\left (
   \frac{1+\sqrt{5}}{2}\right )^5 + \ldots\,
  \ee
represents an unclarified numerical mystery. Why the two quotients
$\varrho_{(A)}$ and $\eta_{(a)}$ should be related at all? And
even if yes, why are they related just to the base $e=
\exp(1)\approx 2.718$ of natural logarithms?

Marginally, we may add a remark that in the next possible study of
the doubly delayed recurrent relations of the above Fibonacci
type,
 \be
 H_{n}=H_{n-3} + H_{n-4}\,
 \label{doubledelay}
 \ee
one could ask why precisely the one-step delay in eq.
(\ref{delay}) should be considered exceptional. The first answer
could be purely pragmatic, stating that algebraic equations
(\ref{latter}) and (\ref{former}) seem to be the only sufficiently
easily manageable pair of definitions of quotients. Indeed, in the
next case with a double-step delay, even the straightforward
asymptotic analysis could be marred by the less transparent
solution of the quartic analogue of eqs. (\ref{latter}) and
(\ref{former}),
 \be
 \tau^4=
 \tau+1\,.
 \label{performer}
 \ee
Its two real roots
 \ben
 -0.72449195900051561159\ldots, \ \ \ \
 1.2207440846057594754\ldots
 \een
and their two complex conjugate partners
 \ben
 \tau_{(\pm)}=
 -0.24812606280262193189\ldots\pm i\,1.0339820609759677567\ldots\,
 \een
can be hardly expressed via reasonably compact formulae. Moreover,
on a deeper level one reveals that the absolute value
$|\tau_{(\pm)}|= 1.0633$ of the two complex roots is bigger than
one. This implies that in the doubly delayed case, the subdominant
component of the asymptotics would not decrease anymore, with all
the possible related complications which were not encountered {\em
just} in the two above-listed cases, viz., in sequences generated
by eqs. (\ref{nedelay}) and (\ref{delay}). This adds a further
background to our belief that only the famous Fibonacci numbers
$F_n$ {\em and} their present one-step-delayed ``Gibonacci"
numbers $G_m$ deserve really an {\em exceptional} attention.

\newpage

\section*{Acknowledgement}

Work partially supported by the grant Nr. A 1048302 of GA AS CR.

\newpage

Table 1. Chains of elements $\spadesuit$ with pairwise confluences
$\heartsuit$ (cf. \cite{Lippschw}).

\hspace{2cm}

$$
\begin{array}{||c||c|c|c|c|c||}
\hline \hline
{\rm }&\multicolumn{5}{c||}{\rm }\\
{\rm length}&\multicolumn{5}{c||}{\rm eligible \ structures}\\
&\multicolumn{5}{c||}{}\\
\hline 1 & \spadesuit &&&&\\
\hline 2 & \spadesuit \spadesuit & \heartsuit &&&\\
\hline 3 & \spadesuit \spadesuit \spadesuit & \spadesuit \heartsuit &
 \heartsuit \spadesuit &&\\
\hline 4 & \spadesuit \spadesuit \spadesuit \spadesuit &
\spadesuit \spadesuit \heartsuit & \spadesuit \heartsuit
\spadesuit & \heartsuit \spadesuit \spadesuit &
\heartsuit \heartsuit \\
\hline 5 & \spadesuit \spadesuit \spadesuit \spadesuit \spadesuit
& \spadesuit \spadesuit \spadesuit \heartsuit & \spadesuit
\spadesuit \heartsuit \spadesuit & \spadesuit
\heartsuit \spadesuit \spadesuit & \heartsuit \spadesuit \spadesuit
 \spadesuit\\
 & \spadesuit \heartsuit \heartsuit & \heartsuit \spadesuit \heartsuit &
\heartsuit \heartsuit \spadesuit &&\\
\hline 6 & \spadesuit \spadesuit \spadesuit \spadesuit \spadesuit
\spadesuit & \spadesuit \spadesuit \spadesuit \spadesuit
\heartsuit & \spadesuit \spadesuit \spadesuit \heartsuit
\spadesuit & \spadesuit \spadesuit
\heartsuit \spadesuit \spadesuit & \spadesuit \heartsuit \spadesuit
 \spadesuit \spadesuit\\
 & \heartsuit \spadesuit \spadesuit \spadesuit \spadesuit & \spadesuit
  \spadesuit \heartsuit \heartsuit &
\spadesuit \heartsuit \spadesuit \heartsuit & \heartsuit
\spadesuit
\spadesuit \heartsuit & \spadesuit \heartsuit \heartsuit \spadesuit \\
 & \heartsuit \spadesuit \heartsuit \spadesuit &  \heartsuit \heartsuit
  \spadesuit
\spadesuit & \heartsuit \heartsuit \heartsuit  & &\\
\hline \hline \ea
$$

\newpage

Table 2. Chains of elements  $\spadesuit$ with confluences in
triplets $\clubsuit$.

\hspace{2cm}

$$
\begin{array}{||c||c|c|c|c|c|c||}
\hline \hline
{\rm }&\multicolumn{6}{c||}{\rm }\\
{\rm length}&\multicolumn{6}{c||}{\rm  eligible \ structures}\\
&\multicolumn{6}{c||}{}\\
\hline 1 & \spadesuit &&&&&\\
\hline 2 & \spadesuit \spadesuit & & &&&\\
\hline 3 & \spadesuit \spadesuit \spadesuit& \clubsuit &&&&\\
\hline 4 & \spadesuit \spadesuit \spadesuit \spadesuit &
\spadesuit \clubsuit  & \clubsuit \spadesuit  &&
 &\\
\hline 5 & \spadesuit \spadesuit \spadesuit \spadesuit \spadesuit
& \spadesuit \spadesuit  \clubsuit &  \spadesuit \clubsuit
\spadesuit &
\clubsuit \spadesuit \spadesuit & &\\
\hline 6 & \spadesuit \spadesuit \spadesuit \spadesuit \spadesuit
\spadesuit & \spadesuit \spadesuit \spadesuit  \clubsuit &
 \spadesuit \spadesuit \clubsuit \spadesuit &
\spadesuit \clubsuit \spadesuit \spadesuit &  \clubsuit \spadesuit
\spadesuit \spadesuit
&\clubsuit \clubsuit\\
\hline 7 & \spadesuit \spadesuit \spadesuit \spadesuit \spadesuit
\spadesuit \spadesuit & \spadesuit \spadesuit \spadesuit
\spadesuit \clubsuit & \spadesuit
 \spadesuit \spadesuit \clubsuit \spadesuit & \spadesuit
\spadesuit \clubsuit \spadesuit \spadesuit & \spadesuit \clubsuit
\spadesuit \spadesuit \spadesuit &
\\
& \clubsuit \spadesuit \spadesuit \spadesuit \spadesuit&
\spadesuit \clubsuit \clubsuit &   \clubsuit \spadesuit
\clubsuit &  \clubsuit \clubsuit \spadesuit &&\\
\hline 8 & \spadesuit \spadesuit  \spadesuit \spadesuit \spadesuit
\spadesuit \spadesuit \spadesuit & \spadesuit \spadesuit
\spadesuit \spadesuit \spadesuit \clubsuit & \spadesuit \spadesuit
 \spadesuit \spadesuit \clubsuit \spadesuit & \spadesuit \spadesuit
\spadesuit \clubsuit \spadesuit \spadesuit & \spadesuit \spadesuit
\clubsuit \spadesuit \spadesuit \spadesuit &
\\
& \spadesuit \clubsuit \spadesuit \spadesuit \spadesuit \spadesuit
&  \clubsuit \spadesuit \spadesuit \spadesuit \spadesuit
\spadesuit & \spadesuit \spadesuit \clubsuit \clubsuit &
\spadesuit \clubsuit \spadesuit
\clubsuit & \clubsuit \spadesuit \spadesuit  \clubsuit  &\\
&    \spadesuit \clubsuit \clubsuit \spadesuit & \clubsuit
\spadesuit \clubsuit \spadesuit & \clubsuit \clubsuit \spadesuit
\spadesuit
 &&&\\
\hline \hline \ea
$$
\newpage

Table 3. Chains of elements  $\spadesuit$ with quadruplets
 $\diamondsuit$.

\hspace{2cm}

$$
\begin{array}{||c||c|c|c|c|c||}
\hline \hline
{\rm }&\multicolumn{5}{c||}{\rm }\\
{\rm length}&\multicolumn{5}{c||}{\rm  eligible \ structures}\\
&\multicolumn{5}{c||}{}\\
\hline 1 & \spadesuit &&&&\\
\hline 2 & \spadesuit \spadesuit & & &&\\
\hline 3 & \spadesuit \spadesuit \spadesuit&  &&&\\
\hline 4 & \spadesuit \spadesuit \spadesuit \spadesuit &
 \diamondsuit  &   &&
 \\
\hline 5 & \spadesuit \spadesuit \spadesuit \spadesuit \spadesuit
&  \spadesuit  \diamondsuit &   \diamondsuit \spadesuit &
& \\
\hline 6 & \spadesuit \spadesuit \spadesuit \spadesuit \spadesuit
\spadesuit & \spadesuit \spadesuit  \diamondsuit &
  \spadesuit \diamondsuit \spadesuit &
 \diamondsuit \spadesuit \spadesuit &
\\
\hline 7 & \spadesuit \spadesuit \spadesuit \spadesuit \spadesuit
\spadesuit \spadesuit &  \spadesuit \spadesuit \spadesuit
\diamondsuit &
 \spadesuit \spadesuit \diamondsuit \spadesuit &
\spadesuit \diamondsuit \spadesuit \spadesuit & \diamondsuit
\spadesuit \spadesuit \spadesuit
\\
\hline 8 & \spadesuit \spadesuit  \spadesuit \spadesuit \spadesuit
\spadesuit \spadesuit \spadesuit &  \spadesuit \spadesuit
\spadesuit \spadesuit \diamondsuit &  \spadesuit
 \spadesuit \spadesuit \diamondsuit \spadesuit &  \spadesuit
\spadesuit \diamondsuit \spadesuit \spadesuit & \spadesuit
\diamondsuit \spadesuit \spadesuit \spadesuit
\\
&  \diamondsuit \spadesuit \spadesuit \spadesuit \spadesuit &
  \diamondsuit \diamondsuit & & &  \\
\hline 9 & \spadesuit \spadesuit  \spadesuit  \spadesuit
\spadesuit \spadesuit \spadesuit \spadesuit \spadesuit &
 \spadesuit \spadesuit \spadesuit \spadesuit \spadesuit \diamondsuit &
\spadesuit \spadesuit
 \spadesuit \spadesuit \diamondsuit \spadesuit &  \spadesuit
  \spadesuit
\spadesuit \diamondsuit \spadesuit \spadesuit &  \spadesuit
\spadesuit \diamondsuit \spadesuit \spadesuit \spadesuit
\\
&   \spadesuit \diamondsuit \spadesuit \spadesuit \spadesuit
\spadesuit & \diamondsuit \spadesuit \spadesuit \spadesuit
\spadesuit \spadesuit & \spadesuit
 \diamondsuit \diamondsuit &  \diamondsuit \spadesuit
\diamondsuit &   \diamondsuit
 \diamondsuit \spadesuit \\
\hline \hline \ea
$$

\end{document}